\documentclass[conf]{new-aiaa}
\usepackage[utf8]{inputenc}
\usepackage{graphicx}
\usepackage{amsmath}
\usepackage[version=4]{mhchem}
\usepackage{siunitx}
\usepackage{multirow}
\usepackage[table,xcdraw]{xcolor}
\usepackage{subfig}
\usepackage{longtable,tabularx}
\usepackage{float}
\title{Vibration Analysis of Timoshenko Beams using Isogeometric Analysis}

\author{Balakrishnan Devarajan\footnote{Department of Biomedical Engineering and Mechanics, dbalak9@vt.edu
}}
\affil{Virginia Polytechnic Institute and State University, Blacksburg, VA 24060, USA}

\begin{document}

\maketitle

\begin{abstract}

In this paper, the finite free-form  beam element is formulated by the isogeometric approach based on the Timoshenko  beam theory to investigate the free vibration behavior of the  beams. The non-uniform rational B-splines (NURBS) functions which define the geometry of the  beam are used as the basis functions for the finite element analysis. In order to enrich the basis functions and to increase the accuracy of the solution fields, the h-, p-, and k-refinement techniques are implemented. The geometry and curvature of the beams are modelled in a unique way based on NURBS. All the effects of the the shear deformation, and the rotary inertia are taken into consideration by the present isogeometric model. Results of the beams for non-dimensional frequencies are compared with other available results in order to show the accuracy and efficiency of the present isogeometric approach. From numerical results, the present element can produce very accurate values of natural frequencies and the mode shapes due to exact definition of the geometry. With higher order basis functions, there is no shear locking phenomenon in very thin beam situations. Finally, the benchmark tests described in this study are provided as future reference solutions for Timoshenko beam vibration problem.

\end{abstract}
\section{Nomenclature}

{\renewcommand\arraystretch{1.0}
\noindent\begin{longtable*}{@{}l @{\quad=\quad} l@{}}

$x,y,z$ & coordinate system defined for the beam\\
$u, v, w$ &	displacements in the $x$, $y$ and $z$ directions respectively\\
$u_0,v_0 $	&	displacements in the $x$ and $y$ directions of the mid plane respectively\\
$\varphi$	&	rotation about the $z$ direction\\
$\Pi$	&	knot vector\\
$B$	&	strain displacement matrix\\
$J$	&   determinant of the jacobian\\
$W_k$	&     corresponding weight of the gauss points\\
$\epsilon_{i}$	&    $i^{\text{th}}$ knot\\
$W(\epsilon)$	&	weighting function\\
$q_i$	&	weight of $i^{\text{th}}$ control point\\
$(P_i)_j$	&	control point\\
$\text{N}^p_i$	&	$i^{\text{th}}$ basis function of order $p$\\
$p$	& 	order of the polynomial\\
$n$ & number of control points\\
$E$ & Young's modulus\\

\end{longtable*}}

\section{Introduction}

The first meaningful observation on the beam vibration appears to be the one conducted by Bernoulli. From his observation, \cite{Li2016} he discovered that the curvature of an elastic beam is proportional to the bending moment. Later, Leonhard Euler accepted Bernoulli’s assumption and made many advances on the elastic curves. In engineering practice, the Euler–Bernoulli’s beam theory has been then most frequently used to predict the natural frequencies of the beams. However, as well-known fact, it overestimates the natural frequencies of the beams especially for higher modes and is eligible for slender beams. Rayleigh [1] introduced rotatory inertia effect and corrected the overestimation of natural frequency for some problems. But in particular problems, the introduction of shear was still required in order to improve the accuracy of natural frequencies. In this context, Timoshenko \cite{khdeir1997buckling} proposed a beam theory which includes the transverse shear deformation and the rotatory inertia effect. Since then, Timoshenko’s beam theory has been widely used in practical engineering problems.

Khdeir and 
Reddy \cite{khdeir1997buckling} derived analytical buckling solutions of refined shear 
deformation beam theories and compared them with classical theories.
Research on the vibration of Timoshenko beams and Mindlin plates can be divided into three categories. Firstly, there exist exact solutions only for a very restricted number of simple cases. Secondly, studies of semi-analytic solutions, including the differential quadrature method, and the boundary characteristic orthogonal polynomials ~\cite{nguyen2011isogeometric, kapoor2013isogeometric} are available. Finally, there are the most widely used discretization methods such as the finite element method and the finite difference method. As it is more useful to have analytical results than to resort to numerical methods, most efforts focus on developing efficient semi-analytic solutions.

Isogeometric analysis (IGA) developed by Hughes et al. \cite{hughes2005isogeometric} can be used 
to implement multiple boundary conditions\cite{devarajan2020thermal,Devarajan2016,devarajan2021analyzing,Miglani2018,devarajan2019thermomechanical} as opposed to other 
approximate energy method formulations. Hence, beam problems with 
multiple support can be tackled using this approach. Moreover, the 
numerical solution obtained by IGA is expected to give 
a better convergence due to its \textit{k} refinement capability as shown by Nguyen \cite{nguyen2011isogeometric}. 
The solution does not oscillate with an increase in the polynomial order as 
opposed to the polynomials used in the conventional finite element polynomial shape 
functions.

As shown in the , a Timoshenko beam is considered. 

Shear effects were considered in the beam formulation, as depicted by Kapoor et al.
~\cite{kapoor2013isogeometric} and Kapania et al. \cite{kapania1989recent} using first order shear deformation theory (Timoshenko 
Beam Theory \cite{timoshenko1922x}) where the displacements are given 
by Eq.~\eqref{sample:equation1} and ~\eqref{sample:equation2}: -

\begin{equation}
\label{sample:equation1}
u(x,y)=u_{0}(x)+y\varphi(x)
\end{equation}
\begin{equation}
\label{sample:equation2}
v(x,y)=v_{0}(x)
\end{equation}

The strain-displacement relationships are given as:-
\begin{equation}
\label{sample:equation3}
\varepsilon_x = \frac{du_0(x)}{dx}+y\frac{d\varphi(x)}{dx}
\end{equation}
\begin{equation}
\gamma_{xy} = \varphi(x)+\frac{dv_0(x)}{dx}
\end{equation}

% The strain, vector corresponding to direct stresses, is given by Eq.~\eqref{sample:equation5} 
% as: -

% \begin{equation}
% \label{sample:equation5}
% \begin{Bmatrix}
% \varepsilon _{x}\\
% \varepsilon _{y}\\
% \gamma _{xy}\\
% \end{Bmatrix}
% =\begin{Bmatrix}
% \frac{du_{0}(x)}{dx}\\
% 0\\
% \varphi +\frac{dv_{0}(x)}{dx}\\
% \end{Bmatrix}
%  +y\begin{Bmatrix}
% \frac{d\varphi(x) }{dx}\\
% 0\\
% 0\\
% \end{Bmatrix}
% \end{equation}

In IGA, mid-plane displacements and rotations can be written as:-
\begin{equation}
\label{sample:equation7}
u_{0}=\sum_{i=1}^{n}{N(\epsilon _{i})u_{i}};\hspace{1cm}
v_{0}=\sum_{i=1}^{n}{N(\epsilon _{i})v_{i}};\hspace{1cm}
\varphi =\sum_{i=1}^{n}{N(\epsilon _{i})\varphi _{i}};\hspace{1cm}n = \text{number\hspace{0.1cm}of\hspace{0.1cm}control\hspace{0.1cm}points}
\end{equation}

Using NURBS basis functions, the in-plane extension, transverse deflection, and the rotations at control points can be expressed as:- 
\begin{equation}
\label{sample:equation8}
\textbf{u}_{0}=\sum_{k=1}^{n}\begin{bmatrix}
N & 0 & 0\\
0 & N & 0\\
0 & 0 & N\\
\end{bmatrix}\\
\begin{bmatrix}
u\\
v\\
\varphi\\
\end{bmatrix}
\end{equation}
Where $\begin{bmatrix}
u & v & \varphi
\end{bmatrix}^T$ are
the degrees of freedom of $\textbf{u}_{0}$ associated with a control point $k$.\\
The strain displacement matrices for bending B$_{f}$, membrane B$_{m}$ and transverse shear B$_{c}$ are:-
\begin{equation}
\label{sample:equation9}
B_{m}=\begin{bmatrix}
\frac{\partial N}{\partial x} & 0 & 0\\
0 & 0 & 0\\
\end{bmatrix}
\end{equation}

\begin{equation}
\label{sample:equation10}
B_{f}=\begin{bmatrix}
0 & 0 & \frac{\partial N}{\partial x}\\
0 & 0 & 0\\
\end{bmatrix}
\end{equation}

\begin{equation}
\label{sample:equation11}
B_{c}=\begin{bmatrix}
0 & 0 & 0\\
0 & \frac{\partial N}{\partial x} & N\\
\end{bmatrix}
\end{equation}

The expression of stiffness matrix is given as:-
\begin{equation}
\label{sample:equation14}
\begin{gathered}
K=b{(\sum_{i=1}^{n}{B_{m}^{T} D
B_{m}}z J W+\frac{1}{2}\sum_{i=1}^{n}{B_{m}^{T} D^ 
B_{f}}z^2 J 
W_{k}}\\+{\frac{1}{2}\sum_{i=1}^{n}{B_{f}^{T} D^
B_{m}}(z^{2}-z^{2}) J 
W}+
{\frac{1}{3}\sum_{i=1}^{n}{B_{f}^{T} D
B_{f}}(z^{3}-z^{3}) J W+\sum_{i=1}^{n}{B_{c}^{T} D_{c}
B_{c}}(z-z)J W)}\\
\end{gathered}
\end{equation}
Where \textit{b} is the breadth of the beam, \textit{J} is the Jacobian and
\begin{equation}
\label{sample:equation13}
D=\begin{bmatrix}
\overline{Q}_{11} & \overline{Q}_{12} & \overline{Q}_{16}\\
\overline{Q}_{12} & \overline{Q}_{22} & \overline{Q}_{26}\\
\overline{Q}_{16} & \overline{Q}_{26} & \overline{Q}_{66}\\
\end{bmatrix},\hspace{0.5cm} {D_c}=\begin{bmatrix}
\overline{Q}_{44} & 0\\
0 & \overline{Q}_{55}\\
\end{bmatrix}\hspace{.5cm}
\end{equation}

The entries of \textbf{${K_{e}}$} (element stiffness matrix) are computed 
using Gauss-Legendre quadrature numerical integration 
technique and the element stiffness matrices are assembled into a global 
stiffness matrix \textbf{$K$}. 

We can compute the mass matrix

\begin{equation}
\label{sample:equation17}
\begin{gathered}
K=\rho\begin{bmatrix}
N & 0 & 0\\
0 & N & 0\\
0 & 0 & N\\
\end{bmatrix}^T
\begin{bmatrix}
h & 0 & 0\\
0 & h & 0\\
0 & 0 & h^3/12\\
\end{bmatrix}
\begin{bmatrix}
N_{k} & 0 & 0\\
0 & N_{k} & 0\\
0 & 0 & N_{k}\\
\end{bmatrix}
\end{gathered}
\end{equation}

% \begin{equation}
% \label{sample:equation15}
% G_{b2}=\begin{bmatrix}
% 0 & \frac{\partial N}{\partial x} & 0\\\
% 0 & 0 & 0\\
% \end{bmatrix}
% \end{equation}

% \begin{equation}
% \label{sample:equation16}
% G_{b3}=\begin{bmatrix}
% 0 & 0 & \frac{\partial N}{\partial x}\\
% 0 & 0 & 0\\
% \end{bmatrix}
% \end{equation}

The non-dimensional natural frequencies are given by $(\widetilde{\omega})$, defined in Eq.~\eqref{sample:equation19} is then evaluated by solving eigenvalue problem 
as.

\begin{equation}
\label{sample:equation18}
\lbrace \lbrack K\rbrack -\omega^2 \lbrack M\rbrack 
\rbrace \lbrack \delta \rbrack =0
\end{equation}

\begin{equation}
\label{sample:equation19}
\widetilde{\omega} =\omega L^2\sqrt{\frac{\rho A}{EI}}
\end{equation}

\section{Solution Approach}
\subsection{Isogeometric Analysis}
\begin{figure}[hbt!]
\centering
\includegraphics[width=0.4\textwidth]{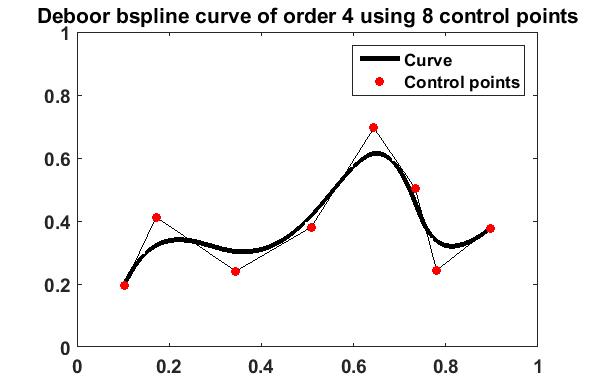}
\caption{A Curve Represented Using NURBS.}
\label{fig:fig2}
\end{figure}

Isogeometric approach (IGA) uses Non-Uniform Rational Basis Spline 
(NURBS) representation as shown in Fig. \ref{fig:fig2} to define both 
geometry and the displacement shape function.

The NURBS basis functions can be defined as

\begin{equation}
\label{sample:equation20}
R_{i}^{p}(\epsilon )=\frac{N_{i,p}(\epsilon 
)q_{i}}{W(\epsilon )}=\frac{N_{i,p}(\epsilon 
)q_{i}}{\sum_{i=1}^{n}{N_{i,p}(\epsilon )}q_{i}(\epsilon )}
\end{equation}

\begin{equation}
\label{sample:equation21}
N_{i,0}(\epsilon )=\bigg\{ 
\begin{gathered}
1,\epsilon _{i}\le \epsilon \le \epsilon _{i+1} \\
0,otherwise \\
\end{gathered}
\end{equation}

\begin{equation}
\label{sample:equation22}
N_{i,p}(\epsilon )=\frac{\epsilon -\epsilon 
_{i}}{\epsilon _{i+p}-\epsilon _{i}}N_{i,p-1}(\epsilon 
)+\frac{\epsilon _{i+p+1}-\epsilon }{\epsilon _{i+p+1}-\epsilon 
_{i+1}}N_{i+1,p-1}(\epsilon )
\end{equation}

Computer Aided Design (CAD) systems are based on spline basis functions, often Non-Uniform Rational B-Splines (NURBS). B-Splines are widely used in research and industry to parameterize complex geometry. A particular example is research by De et al. where B-splines have been used to parameterize unconventional concept models of aircrafts and vehicles ~\cite{de2019structural, jrad2017global, robinson2016aeroelastic, de2021lightweight, de2017sparibs, de2019unconventional, de2017structural, de2018structural, de2020algorithms, demanual}.
It has to be noted that the analysis in the current work is restricted to straight beams and hence B-splines were used \textit{i.e.}  the weights associated with all the control points are set to 1.  
In FEM, elements are used to divide discretize geometry; whereas, in IGA patches are used. In finite element analysis (FEA), the parameter space is the reference element 
which is mapped into each element in the physical space. However in IGA, parameter space is the whole patch which is further 
divided into elements by a knot vector in each direction as given in Eq.~\eqref{sample:equation23} 
\begin{equation}
\label{sample:equation23}
\Pi ={\epsilon _{1},\epsilon _{2},...,\epsilon 
_{i},...,\epsilon _{n+p+1}}
\end{equation}

where $\epsilon _{1},\epsilon _{2},...,\epsilon 
_{i},...,\epsilon _{n+p+1}$ are the knot vectors.

The length of the knot vector is given by 
\begin{equation}
\label{sample:equation24}
\Pi=n+p+1.
\end{equation}

\subsection{Refinement}
The refinement in IGA analysis is of three types. The first type is the order elevation. Analogous to \textit{p} refinement in FEA, the order elevation involves increasing the multiplicity of each knot value by one. The geometry and the parameterization of the physical curve are not changed. However, the number of basis functions and control points increase.
    Next is knot insertion. It has similarities to \textit{h} refinement in FEA, which involves splitting the elements. In IGA, knot insertion involves adding new knot values between the existing knots. This creates more elements, since the elements are bounded by knots of different values.
The last and most widely used refinement is the \textit{k} refinement. This has no analog in FEA and is one of the factors that make IGA analysis very powerful. The \textit{k} refinement is a combination of order elevation and knot insertion. This is done by increasing just the multiplicity of the first and the last knot values.  The continuity across knot values is then increased by the same number as the number of multiplicities in order elevation (i.e. a $C^{\text{p-1}}$ continuity). The \textit{k} refinement is widely used and possesses improved computational efficiency due to lesser number of basis functions. Mapping from reference to parametric and parametric to physical space can be found in work by Nguyen \textit{et al.} \cite{nguyen2015isogeometric}.
\section{Results and Discussion}

\begin{table}[H]
\centering
\caption{Non-dimensionalized frequency parameter  of the Timoshenko beam (pinned–pinned boundary condition, nu=0.3, alpha=5/6}
\label{tab:ls_compositenostiff}
\begin{tabular}{|c|c|c|c|c|c|c|c|c|}
\hline
{  }                                & {  }                               & \multicolumn{7}{c|}{{  }}                         \\
  {{  \textbf{Mode}}} &   {{  \textbf{CLT}}} & \multicolumn{7}{c|}{  {{  \textit{\textbf{h/L}}}}}                                                                                                                                                                                                                                                                                                                          \\ \hline
{  }                                & {  }                               & {  }                                 & {  }                                 & {  }                                & {  }                                & {  }                                & {  }                               & {  }                               \\
  {{  \textbf{}}}     &   {{  \textbf{}}}    &   {{  \textbf{0.002}}} &   {{  \textbf{0.005}}} &   {{  \textbf{0.01}}} &   {{  \textbf{0.02}}} &   {{  \textbf{0.05}}} &   {{  \textbf{0.1}}} &   {{  \textbf{0.2}}} \\ \hline
{  }                                & {  }                               & {  }                                 & {  }                                 & {  }                                & {  }                                & {  }                                & {  }                               & {  }                               \\
  {{  1}}             &   {{  3.14159}}      &   {{  3.1417}}         &   {{  3.1415}}         &   {{  3.1413}}        &   {{  3.1405}}        &   {{  3.135}}         &   {{  3.1157}}       &   {{  3.0453}}       \\ \hline
{  }                                & {  }                               & {  }                                 & {  }                                 & {  }                                & {  }                                & {  }                                & {  }                               & {  }                               \\
  {{  2}}             &   {{  6.28319}}      &   {{  6.2839}}         &   {{  6.2828}}         &   {{  6.2811}}        &   {{  6.2747}}        &   {{  6.2314}}        &   {{  6.0907}}       &   {{  5.6716}}       \\ \hline
{  }                                & {  }                               & {  }                                 & {  }                                 & {  }                                & {  }                                & {  }                                & {  }                               & {  }                               \\
  {{  3}}             &   {{  9.42478}}      &   {{  9.4271}}         &   {{  9.4234}}         &   {{  9.4177}}        &   {{  9.3964}}        &   {{  9.2554}}        &   {{  8.8405}}       &   {{  7.8395}}       \\ \hline
{  }                                & {  }                               & {  }                                 & {  }                                 & {  }                                & {  }                                & {  }                                & {  }                               & {  }                               \\
  {{  4}}             &   {{  12.5664}}      &   {{  12.5718}}        &   {{  12.5632}}        &   {{  12.5497}}       &   {{  12.4995}}       &   {{  12.1814}}       &   {{  11.3431}}      &   {{  9.6571}}       \\ \hline
{  }                                & {  }                               & {  }                                 & {  }                                 & {  }                                & {  }                                & {  }                                & {  }                               & {  }                               \\
  {{  5}}             &   {{  15.708}}       &   {{  15.7187}}        &   {{  15.7017}}        &   {{  15.6755}}       &   {{  15.5787}}       &   {{  14.9928}}       &   {{  13.6132}}      &   {{  11.2221}}      \\ \hline
{  }                                & {  }                               & {  }                                 & {  }                                 & {  }                                & {  }                                & {  }                                & {  }                               & {  }                               \\
  {{  6}}             &   {{  18.8496}}      &   {{  18.8682}}        &   {{  18.8388}}        &   {{  18.7937}}       &   {{  18.6286}}       &   {{  17.6812}}       &   {{  15.6792}}      &   {{  12.6023}}      \\ \hline
{  }                                & {  }                               & {  }                                 & {  }                                 & {  }                                & {  }                                & {  }                                & {  }                               & {  }                               \\
  {{  7}}             &   {{  21.9911}}      &   {{  22.021}}         &   {{  21.9742}}        &   {{  21.9028}}       &   {{  21.645}}        &   {{  20.245}}        &   {{  17.5707}}      &   {{  13.0323}}      \\ \hline
{  }                                & {  }                               & {  }                                 & {  }                                 & {  }                                & {  }                                & {  }                                & {  }                               & {  }                               \\
  {{  8}}             &   {{  25.1327}}      &   {{  25.1776}}        &   {{  25.1075}}        &   {{  25.0014}}       &   {{  24.6237}}       &   {{  22.6866}}       &   {{  19.3144}}      &   {{  13.4443}}      \\ \hline
{  }                                & {  }                               & {  }                                 & {  }                                 & {  }                                & {  }                                & {  }                                & {  }                               & {  }                               \\
  {{  9}}             &   {{  28.2743}}      &   {{  28.3388}}        &   {{  28.2386}}        &   {{  28.0882}}       &   {{  27.5614}}       &   {{  25.0117}}       &   {{  20.9328}}      &   {{  13.8434}}      \\ \hline
{  }                                & {  }                               & {  }                                 & {  }                                 & {  }                                & {  }                                & {  }                                & {  }                               & {  }                               \\
  {{  10}}            &   {{  31.4159}}      &   {{  31.5052}}        &   {{  31.3672}}        &   {{  31.162}}        &   {{  30.4553}}       &   {{  27.2271}}       &   {{  22.4445}}      &   {{  14.4378}}      \\ \hline
\end{tabular}
\end{table}
\begin{table}[H]
\centering
\caption{Non-dimensionalized frequency parameter of the Timoshenko beam (clamped–clamped boundary condition, nu=0.3, alpha=5/6}
\label{tab:my-table}
\begin{tabular}{|c|c|c|c|c|c|c|c|c|}
\hline
{  \textbf{Mode}} & {  \textbf{CLT}}                 & \multicolumn{7}{c|}{{  \textit{\textbf{h/L}}}}                                                                                                                                                                                                                                                                                                                       \\ \hline
{  \textbf{}}     & {  \textbf{}}                    & {  \textbf{0.002}}               & {  \textbf{0.005}}               & {  \textbf{0.01}}                & {  \textbf{0.02}}                & {  \textbf{0.05}}                & {  \textbf{0.1}}                 & {  \textbf{0.2}}                 \\ \hline
1                                    & \multicolumn{1}{l|}{{  4.73004}} & \multicolumn{1}{l|}{{  4.72998}} & \multicolumn{1}{l|}{{  4.72963}} & \multicolumn{1}{l|}{{  4.72840}} & \multicolumn{1}{l|}{{  4.72350}} & \multicolumn{1}{l|}{{  4.68991}} & \multicolumn{1}{l|}{{  4.57955}} & \multicolumn{1}{l|}{{  4.24201}} \\ \hline
2                                    & {  7.8532}                       & 7.9272                                              & 7.8877                                              & 7.8606                                              & 7.8321                                              & 7.7042                                              & 7.3314                                              & 6.418                                               \\ \hline
3                                    & {  10.9956}                      & 11.1019                                             & 11.0423                                             & 10.9991                                             & 10.9396                                             & 10.641                                              & 9.8563                                              & 8.2853                                              \\ \hline
4                                    & {  14.1372}                      & 14.2781                                             & 14.1946                                             & 14.1304                                             & 14.0223                                             & 13.4622                                             & 12.1456                                             & 9.9038                                              \\ \hline
5                                    & {  17.2788}                      & 17.4574                                             & 17.3452                                             & 17.2541                                             & 17.0761                                             & 16.1602                                             & 14.2327                                             & 11.3488                                             \\ \hline
6                                    & {  20.4204}                      & 20.6401                                             & 20.4939                                             & 20.3685                                             & 20.0964                                             & 18.7332                                             & 16.149                                              & 12.6403                                             \\ \hline
7                                    & {  23.5619}                      & 23.827                                              & 23.6404                                             & 23.4724                                             & 23.0791                                             & 21.184                                              & 17.9218                                             & 13.4567                                             \\ \hline
8                                    & {  26.7035}                      & 27.0187                                             & 26.7843                                             & 26.5644                                             & 26.0209                                             & 23.5185                                             & 19.5727                                             & 13.8102                                             \\ \hline
9                                    & {  29.8451}                      & 30.216                                              & 29.9254                                             & 29.6431                                             & 28.9188                                             & 25.7439                                             & 21.1189                                             & 14.4806                                             \\ \hline
10                                   & {  32.9867}                      & 33.4195                                             & 33.0636                                             & 32.7075                                             & 31.7707                                             & 27.8682                                             & 22.5739                                             & 14.9384                                             \\ \hline
\end{tabular}
\end{table}

\section{Conclusions and Future Work}

A isogeometric analysis using NURBS basis functions is applied to the free vibration analysis of Timoshenko beams. Because the formulation is so simple and efficient, allowing the process of calculating weighting coefficients and characteristic polynomials to be avoided, this method has merits over other semi-analytic methods. Rapid convergence, good accuracy as well as the conceptual simplicity characterize the isogeometric analysis method. The results from this method agree with those of Bernoulli–Euler beams  when the thickness-to-length (radius) ratio is very small, however, deviate considerably as the thickness-to-length (radius) ratio grows larger.  The isogeometric analysis a computationally efficient method that could be used along with powerful modern optimization algorithms like Particle Swarm Optimization and Differential Evolution ~\cite{biswas2021improving, saha2022chagskode, saha2022framework} to perform structural optimization and come up parameters for with light-weight designs.
\bibliography{isogeometric}
\end{document}